\documentclass[aps,showpacs,12pt,groupedaddress,superscriptaddress]{revtex4-1}

\usepackage[    bookmarks,
                 bookmarksopen = true,
                 bookmarksnumbered = true,
                 linktocpage,
                 colorlinks = true,
                 linkcolor = blue,
                 urlcolor  = blue,
                 citecolor = blue,
                 anchorcolor = green,
                 hyperindex = true,
                 hyperfigures]
                 {hyperref}

\usepackage{multirow}

\usepackage{graphicx,xcolor}

\begin{document}

\preprint{This line only printed with preprint option}

\title{An equation-of-state-meter of QCD transition from deep learning}

\author{Long-Gang Pang*}
\affiliation{Frankfurt Institute for Advanced Studies, 60438 Frankfurt am Main, Germany}
\affiliation{Department of Physics, University of California, Berkeley, CA 94720, USA}
\affiliation{Nuclear Science Division, Lawrence Berkeley National Laboratory, Berkeley, CA 94720, USA}

\author{Kai Zhou*}
\affiliation{Frankfurt Institute for Advanced Studies, 60438 Frankfurt am Main, Germany}
\affiliation{Institut f\"ur Theoretische Physik, Goethe Universit\"at, 60438 Frankfurt am Main, Germany}

\author{Nan Su*}
\affiliation{Frankfurt Institute for Advanced Studies, 60438 Frankfurt am Main, Germany}

\author{Hannah Petersen}
\affiliation{Frankfurt Institute for Advanced Studies, 60438 Frankfurt am Main, Germany}
\affiliation{Institut f\"ur Theoretische Physik, Goethe Universit\"at, 60438 Frankfurt am Main, Germany}
\affiliation{GSI Helmholtzzentrum f\"ur Schwerionenforschung, 64291 Darmstadt, Germany}

\author{Horst St\"ocker}
\affiliation{Frankfurt Institute for Advanced Studies, 60438 Frankfurt am Main, Germany}
\affiliation{Institut f\"ur Theoretische Physik, Goethe Universit\"at, 60438 Frankfurt am Main, Germany}
\affiliation{GSI Helmholtzzentrum f\"ur Schwerionenforschung, 64291 Darmstadt, Germany}

\author{Xin-Nian Wang}
\affiliation{Key Laboratory of Quark and Lepton Physics (MOE) and Institute of Particle Physics, Central China Normal University, Wuhan, 430079, China}
\affiliation{Nuclear Science Division, Lawrence Berkeley National Laboratory, Berkeley, CA 94720, USA}

\begin{abstract}

Supervised learning with a deep convolutional neural network is used to identify the  QCD equation of state (EoS) employed in relativistic hydrodynamic simulations of heavy-ion collisions from the simulated final-state particle spectra $\rho(p_T,\Phi)$.
High-level correlations of $\rho(p_T,\Phi)$ learned by the neural network act as an effective ``EoS-meter'' in detecting the nature of the QCD transition.  
The EoS-meter is model independent and insensitive to other simulation inputs, especially the initial conditions.
Thus it provides a powerful direct-connection of heavy-ion collision observables with the bulk properties of QCD.

\end{abstract}



\maketitle

\section{Introduction}

Deep learning (DL) is a branch of machine learning that learns multiple levels of representations from data~\cite{dl1,dl2}. DL has been successfully applied in pattern recognition and classification tasks such as image recognition and language processing. Recently, the application of DL to physics research is rapidly growing, such as in particle physics~\cite{Baldi:2014kfa,Baldi:2014pta,Searcy:2015apa,Barnard:2016qma,Moult:2016cvt}, nuclear physics~\cite{Utama:2016tcl}, and condensed matter physics~\cite{rg&dl,ml-phases,quantum-many-body-ann,eos-bm,sign,str-f}. DL is shown to be very powerful in extracting pertinent features especially for complex non-linear systems with high-order correlations that conventional techniques are unable to tackle. This suggests that it could be utilized to unveil hidden information from the highly implicit data of heavy-ion experiments.

\begin{figure}[htb!]
\centering
\includegraphics[width=0.6\textwidth]{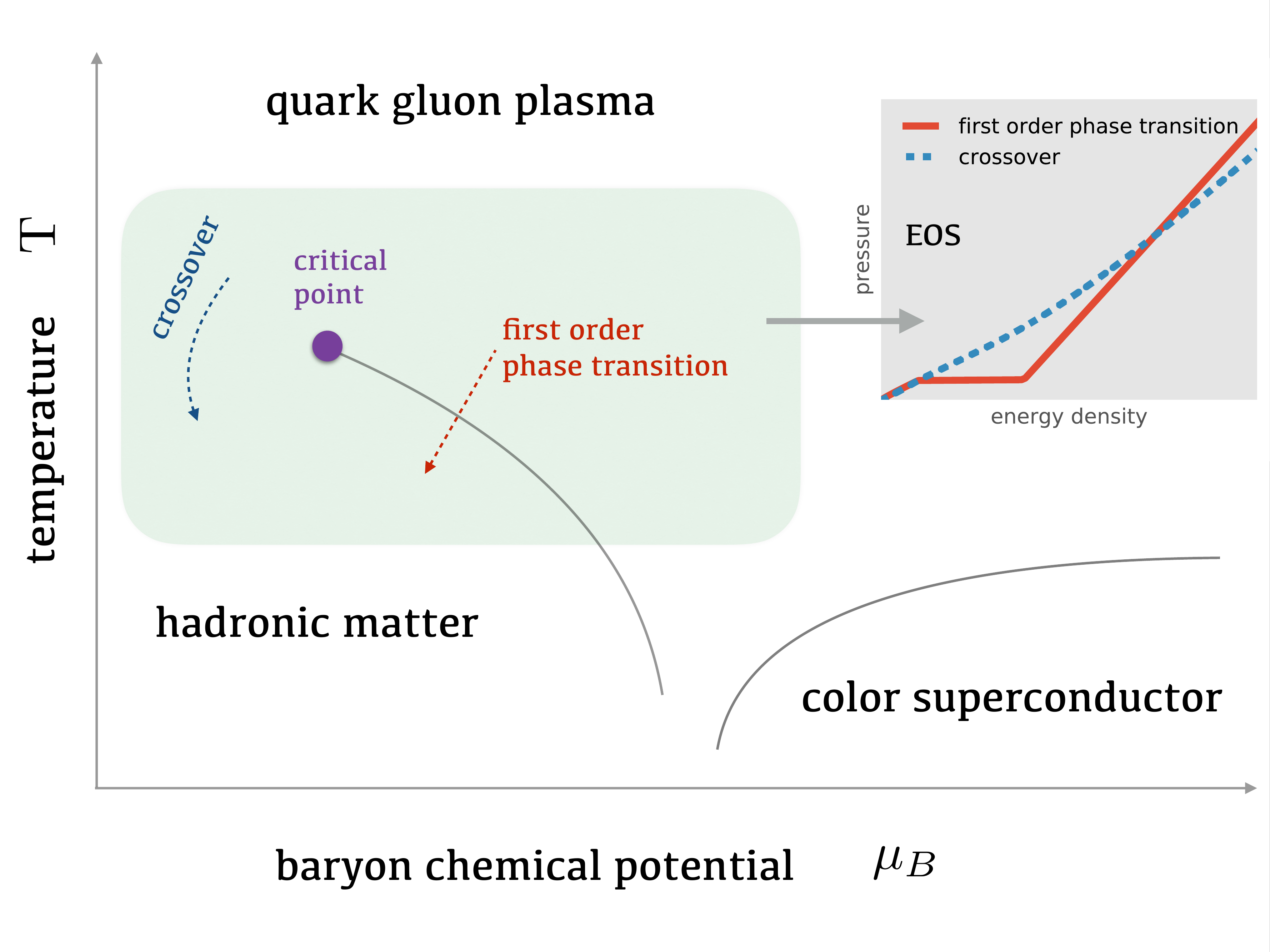}
\caption{\label{fig:phase} Conjectured QCD phase diagram and equations of state for the crossover and the first order phase transition.
}
\end{figure}

Strong interaction in nuclear matter is governed by the theory of Quantum Chromodynamics (QCD). It predicts a transition from the normal nuclear matter, in which the more fundamental constituents, quarks and gluons, are confined within the domains of nucleons, to a new form of matter with freely roaming quarks and gluons as one increases the temperature or density. The QCD transition is conjectured to be a crossover at small density (and moderately high temperature), and first order at moderate density (and lower temperature), with a critical point separating the two, see Fig.~\ref{fig:phase} for a schematic QCD phase diagram and \cite{Stoecker:1986ci,Stephanov:2007fk,Fukushima:2010bq} for some reviews. One primary goal of ultra-relativistic heavy-ion collisions is to study the QCD transition.
Though it is believed that strongly coupled QCD matter can be formed in heavy-ion collisions at the Relativistic Heavy Ion Collider (RHIC, Brookhaven National Laboratory, USA), Large Hadron Collider (LHC, European Organization for Nuclear Research, Switzerland), and at the forthcoming Facility for Anti-proton and Ion Research (FAIR, GSI Helmholtz Centre for Heavy Ion Research, Germany), a direct access to the bulk properties of the matter such as the equation of state (EoS) and transport coefficients is impossible due to the highly dynamical nature of the collisions. In heavy-ion collisions where two high-energy nuclei collide along the longitudinal ($z$) direction,  what experiments measure directly are the final-state particle distributions in longitudinal momentum (rapidity), transverse momentum $p_T$ and azimuthal angle $\Phi$. 
Current efforts to extract physical properties of the QCD matter from experimental data are through direct comparisons with model
calculations of event-averaged and predefined observables such as anisotropic flow \cite{Luzum:2008cw} or global fitting of a set of observables with Bayesian method \cite{Pratt:2015zsa,Bernhard:2016tnd}. However, event-by-event raw data on $\rho(p_T,\Phi)$ at different rapidities provide much more information that contains hidden correlations. These hidden correlations can be sensitive to physical properties of the system but independent of other model parameters.

The aim of the present exploratory study is a first step in directly connecting QCD bulk properties and raw data of heavy-ion collisions using state-of-the-art deep-learning techniques. 
We use the relativistic hydrodynamic model which has been very successful
in simulating heavy-ion collisions and connecting experiments with theory~\cite{Heinz:2009xj,Romatschke:2009im,Teaney:2009qa,Gale:2013da,Strickland:2014pga}. We find unique encoders of bulk properties (here we focus on the EoS) inside $\rho(p_T,\Phi)$ in terms of high-level representations using deep-learning techniques, which are not captured by conventional observables. This is achieved by constructing a convolutional neural network (CNN) and training it with labeled $\rho(p_T,\phi)$ of charged pions generated from the relativistic hydrodynamic program CLVisc~\cite{Pang:2012he,Pang:2014ipa} with two different EoSs as input: crossover~\cite{Huovinen:2009yb} and first order~\cite{Sollfrank:1996hd}. The CNN is then trained with supervision in identifying different EoSs. The performance is surprisingly robust against other simulation parameters such as the initial conditions, equilibrium time $\tau_0$, transport coefficients and freeze out temperature. The supervised learning with deep CNN identifies the hydrodynamic response which is much more tolerant to uncertainties in the initial conditions. $\rho(p_T,\phi)$ as generated by independent simulations (CLVisc with different setup parameters and another hydrodynamic package iEBE-VISHNU~\cite{Shen:2014vra} which implements a different numerical solver for partial differential equations) are used for testing -- on average a larger than $95\%$ testing accuracy is obtained. It has been recently pointed out that model-dependent features (features in the training data that depends on the simulation model and parameters) may generate large uncertainties in the network performance~\cite{Barnard:2016qma}. The network we develop below is, however, not sensitive to these model-dependent features.

\section{Results and discussions}

The evolution of strongly coupled QCD matter can be well described by second-order dissipative hydrodynamics governed by $\partial_{\mu} T^{\mu \nu} = 0$, with $T^{\mu \nu}$ the energy-momentum tensor containing viscous corrections governed by the Israel-Stewart equations~\cite{Heinz:2009xj,Romatschke:2009im}. In order to close the hydrodynamic equations, one must supply the EoS of the medium as one crucial input. The nature of the QCD transition in the EoS strongly affects the hydrodynamic evolution~\cite{Stoecker:2004qu}, since different transitions are associated with different pressure gradients which consequently induce different expansion rates, see the small chart in Fig.~1. Final $\rho(p_T,\Phi)$ are obtained from the Cooper-Frye formula for particle $i$ at mid-rapidity
\begin{equation}
\rho(p_T,\Phi) \equiv  \frac{dN_i}{dY p_T dp_T d\Phi} = g_i \int_{\sigma} p^{\mu} d\sigma_{\mu} f_i \,,
\end{equation}
Here $N_i$ is the particle number density, $Y$ is the rapidity, $g_i$ is the degeneracy, $d\sigma_{\mu}$ is the freeze-out hypersurface element, $f_i$ is the thermal distribution. In the following, we employ the lattice-EoS parametrization~\cite{Huovinen:2009yb} (dubbed as EOSL) for the crossover transition and Maxwell construction~\cite{Sollfrank:1996hd} (dubbed as EOSQ) for the first-order phase transition.

\subsection{Training and testing datasets}

The training dataset of $\rho(p_T,\Phi)$ (labelled with EOSL or EOSQ) is generated by event-by-event hydrodynamic package CLVisc~\cite{Pang:2012he,Pang:2014ipa} with fluctuating AMPT initial conditions~\cite{Lin:2004en}. The simulation generated about 22000 $\rho(p_T,\Phi)$ for different types of collisions. Then the size of the training dataset is doubled by label-preserving left-right flipping along the $\Phi$ direction. In Tab.~\ref{tab:training} we list the details of the training dataset.

\begin{table}[htp]
\begin{center}
\begin{tabular}{cc|c|c|c|c|}
\cline{1-6}
\multicolumn{2}{|c|}{TRAINING}  & \multicolumn{2}{|c|}{$\eta/s=0$}  &  \multicolumn{2}{|c|}{$\eta/s=0.08$} \\ 
\cline{3-6} 
\multicolumn{2}{|c|}{DATASET}  & EOSL & EOSQ & EOSL & EOSQ \\ 
\cline{1-6}
\multicolumn{2}{|c|}{Au-Au $\sqrt{s_{NN}}=200\,$GeV} & 7435 & 5328 & 500 & 500 \\  
\cline{1-6}
\multicolumn{2}{|c|}{Pb-Pb $\sqrt{s_{NN}}=2.76\,$TeV} & 4967 & 2828 & 500 & 500 \\ \hline
\end{tabular}
\end{center}
\caption{Training dataset: numbers of $\rho(p_T,\Phi)$ generated by the CLVisc hydrodynamic package with the AMPT initial conditions in the centrality range $0-60\%$. $\eta/s$ is ratio of shear viscosity to entropy density. $\tau_0=0.4\,$fm for the Au-Au collisions and $\tau_0=0.2\,$fm for the Pb-Pb collisions. The freeze-out temperature is set to be 137\,MeV.}
\label{tab:training}
\end{table}

The testing dataset contains two groups of samples. In the first group, we generate 7343 $\rho(p_T,\Phi)$  events using the second-order event-by-event hydrodynamic package iEBE-VISHNU~\cite{Shen:2014vra} with MC-Glauber initial condition. In the second group, we generate 8917 $\rho(p_T,\Phi)$ events using the CLVisc package with the IP-Glasma-like initial condition~\cite{Gale:2012rq,Bernhard:2016tnd}. The testing datasets are constructed to explore very different regions of parameters as compared to training dataset. The details are listed in Tab.~\ref{tab:testing}. Note that all the training and testing $\rho(p_T,\Phi)$ are preprocessed by $\rho' = \rho / \rho_{max} - 0.5$ to normalize the input data.

\begin{table}[htp]
\begin{center}
\begin{tabular}{cc|c|c|c|c|c|c|}
\cline{1-8}
\multicolumn{8}{|c|}{TESTING DATASET GROUP 1 : iEBE-VISHNU + MC-Glauber} \\
\cline{1-8}
\multicolumn{2}{|c|}{Centrality:}  & \multicolumn{2}{|c|}{$\eta/s\in [0, 0.05]$}  & \multicolumn{2}{|c|}{$\eta/s \in (0.05, 0.10]$}  &  \multicolumn{2}{|c|}{$\eta/s=(0.10, 0.16]$} \\
\cline{3-8}
\multicolumn{2}{|c|}{10-60\%}  & EOSL & EOSQ & EOSL & EOSQ & EOSL & EOSQ \\
\cline{1-8}
\multicolumn{2}{|c|}{Au-Au $\sqrt{s_{NN}}=200$ GeV} & 650 & 850 & 900 & 750 & 200 & 950 \\  
\cline{1-8}
\multicolumn{2}{|c|}{Pb-Pb $\sqrt{s_{NN}}=2.76$ TeV} & 500 & 650 & 600 & 644 & 499 & 150 \\  
\cline{1-8}
\multicolumn{8}{|c|}{TESTING DATASET GROUP 2 : CLVisc + IP-Glasma} \\
\cline{1-8}
\multicolumn{2}{|c|}{Au-Au $\sqrt{s_{NN}}=200$ GeV}  & \multicolumn{3}{|c|}{EOSL}  & \multicolumn{3}{|c|}{EOSQ}  \\
\cline{3-8}
\multicolumn{2}{|c|}{b$\lesssim$8 fm \& $\eta/s=0$} & \multicolumn{3}{|c|}{4165}  & \multicolumn{3}{|c|}{4752} \\ \hline
\end{tabular}
\end{center}
\caption{Testing dataset: numbers of $\rho(p_T,\Phi)$ generated by the CLVisc and iEBE-VISHNU hydrodynamic packages with different initial conditions. $\eta/s$ is ratio of shear viscosity and entropy density. $b$ is the impact parameter. $\tau_0=0.6\,$fm for all the collisions. In iEBE-VISHNU simulations, the freeze-out temperature is varied in the range $[115, 142]\,$MeV. In CLVisc simulations, the freeze-out temperature is set to be 137\,MeV.}
\label{tab:testing}
\end{table}

\subsection{The existence of physical encoders and  neural-network decoder}

\begin{table}[htp]
\begin{center}
\begin{tabular}{cc|c|c|c|}
\cline{1-5}
\multicolumn{2}{|c|}{TESTING DATA} & GROUP 0 & GROUP 1  & GROUP 2 \\   
\cline{1-5}
\multicolumn{2}{|c|}{Number of events} & 4000 & 7343 & 8916 \\  
\cline{1-5}
\multicolumn{2}{|c|}{Accuracy} & $99.88 \pm 0.04\%$ & $93.46 \pm 1.35\%$  & $95.12 \pm 3.08\%$   \\ 
\hline
\end{tabular}
\end{center}
\caption{Testing accuracies for three groups (CLVisc with AMPT initial condition, iEBE-VISHNU and CLVisc with the IP-Glasma-like initial condition) of the testing dataset.}
\label{tab:results}
\end{table}%
After training and validating the network, it is tested on the testing dataset of $\rho(p_T,\Phi)$ events (see Sec.~\ref{s3} for the details of our neural-network model). As shown in Tab.~\ref{tab:results}, high prediction accuracies -- on average larger than $95\%$ with small model uncertainties given by a 10-fold cross validation tests -- are achieved for these three groups of testing datasets, which indicates that our method is highly independent of initial conditions. The network is robust against shear viscosity and $\tau_0$ due to the inclusion of events with different $\eta/s$ and $\tau_0$ in the training. In the testing stage the neural network identifies the type of the QCD transition solely from the spectra of each single event. Furthermore, in the training only one freeze-out temperature is used, while the network is tolerant to a wide range of freeze-out temperatures during the testing. For simplicity, the exploratory study has not included pions from resonance decays (the hadronic transport module UrQMD is switched off in iEBE-VISHNU to exclude contributions from resonance decays in testing data).

For complex and dynamically evolving systems, the final states may not contain enough information to retrieve the physical properties of initial and intermediate states due to entropy production (information loss) during the evolution. The mean prediction accuracy decreases from $97.1\%$ (for $\eta/s=0.0$) to $96.6\%$ (for $\eta/s=0.08$) and $87\%$ (for $\eta/s=0.16$) in the 10-fold cross validation for testing GROUP 1. Besides, the construction of conventional observables may introduce further information loss due to projection of raw data to lower dimensions, as well as information interference due to its sensitivity to multiple factors. These make it yet unclear how to reliably extract physical properties from raw data. Our study firmly demonstrates how to detect the existence of physical encoders in final states with deep CNN decoders, and sets the stage for further applications, such as identifying all relevant physical properties of the systems.

\subsection{Observation from the neural-network decoder}

In order to get physical insights from the neural-network model, it is instructive to visualize the complex dependences learned by the network. For this purpose, we employ the recently developed Prediction Difference Analysis method~\cite{robnik2008explaining,Luisa:2017}. This method uses the observation that replacing one feature in the input image can induce a sizable prediction difference if that feature is important for classification decision. The prediction differences can be visualized as the importance maps of all the input features for the classification network.
\begin{figure}[!htp]
  \centering
    \includegraphics[width=0.33\textwidth, trim={0.5cm 0cm 0.5cm 0},clip]{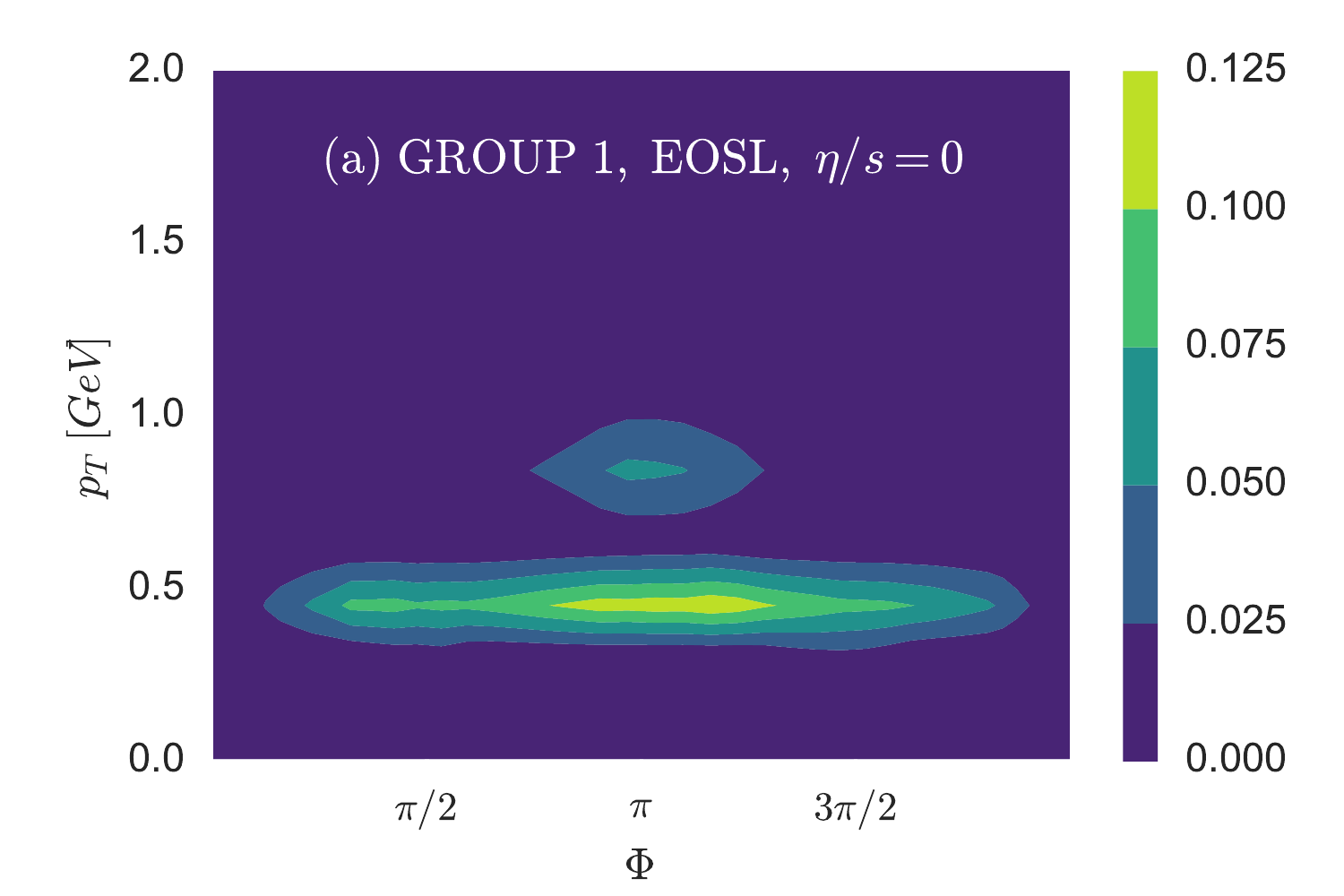}\includegraphics[width=0.33\textwidth, trim={0.5cm 0cm 0.5cm 0},clip]{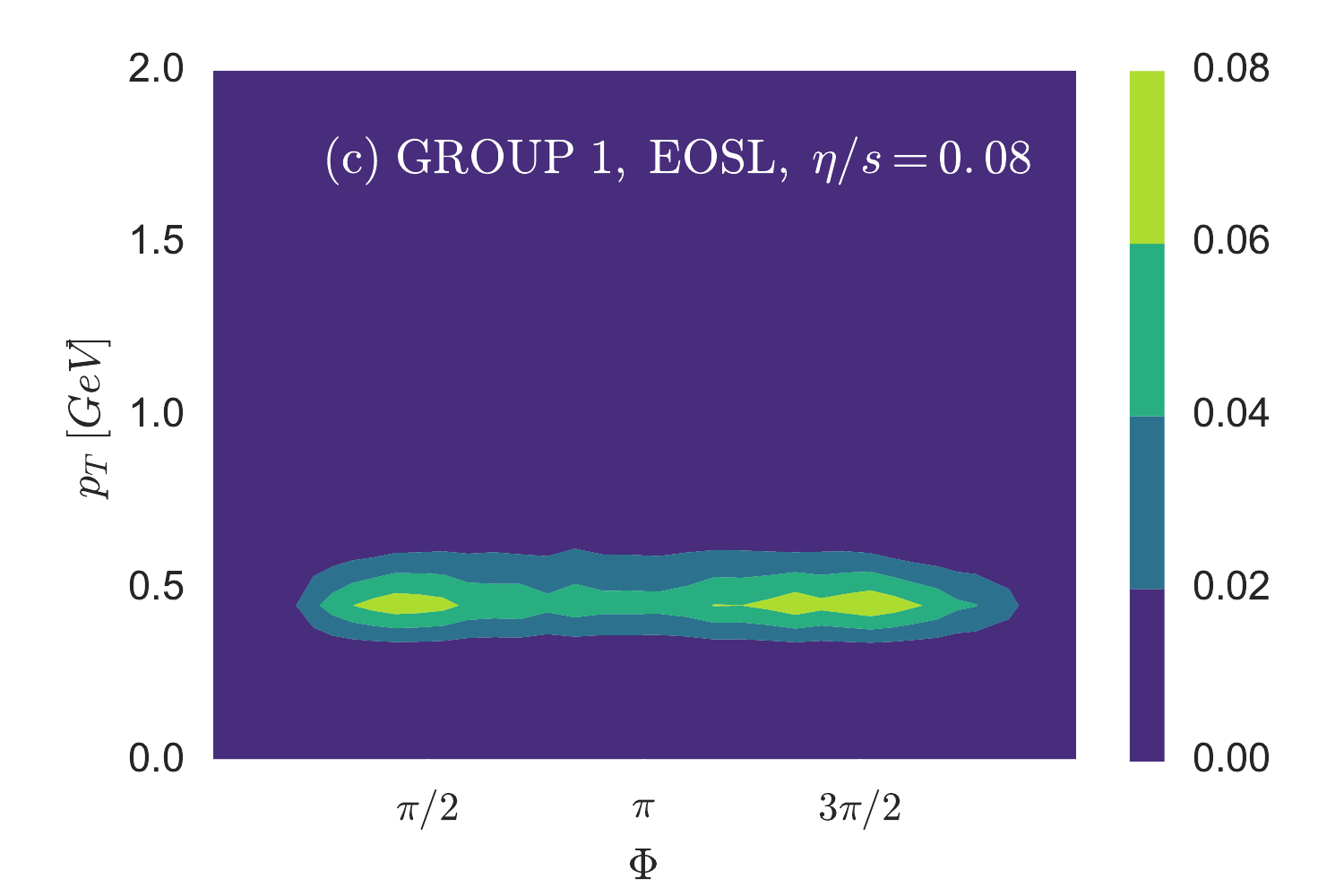}\includegraphics[width=0.33\textwidth, trim={0.5cm 0cm 0.5cm 0},clip]{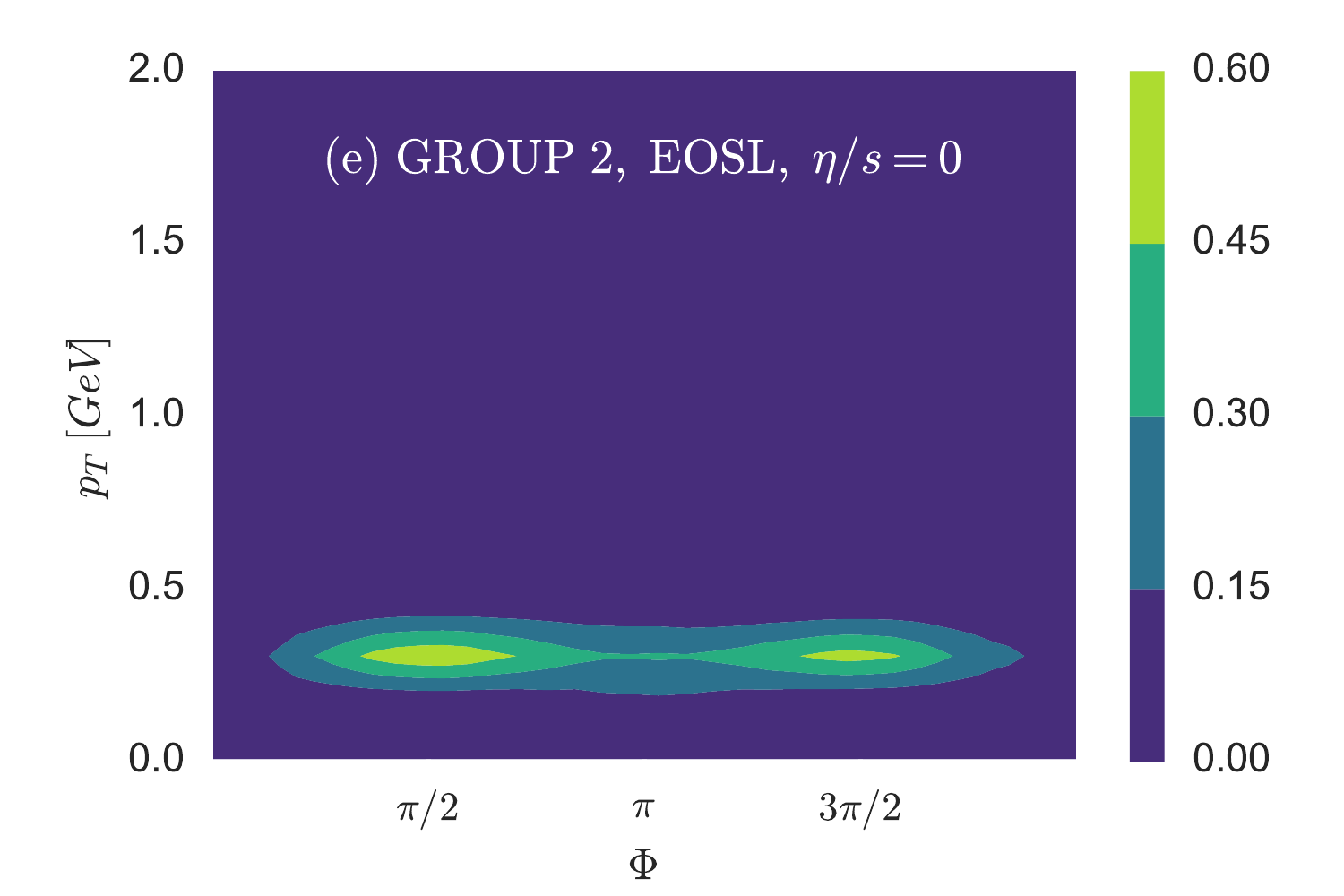}\
    \includegraphics[width=0.33\textwidth, trim={0.5cm 0cm 0.5cm 0},clip]{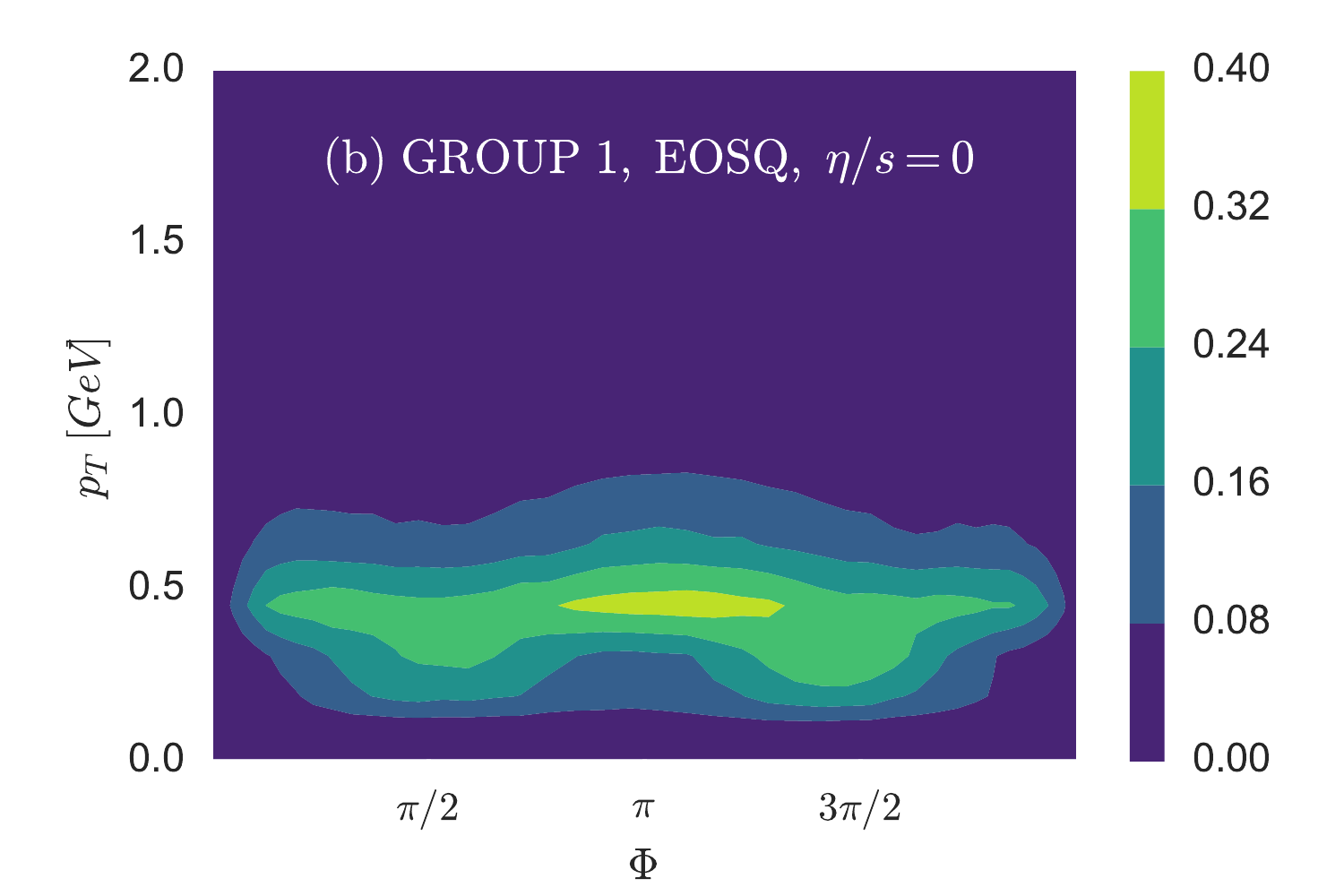}   \includegraphics[width=0.33\textwidth, trim={0.5cm 0cm 0.5cm 0},clip]{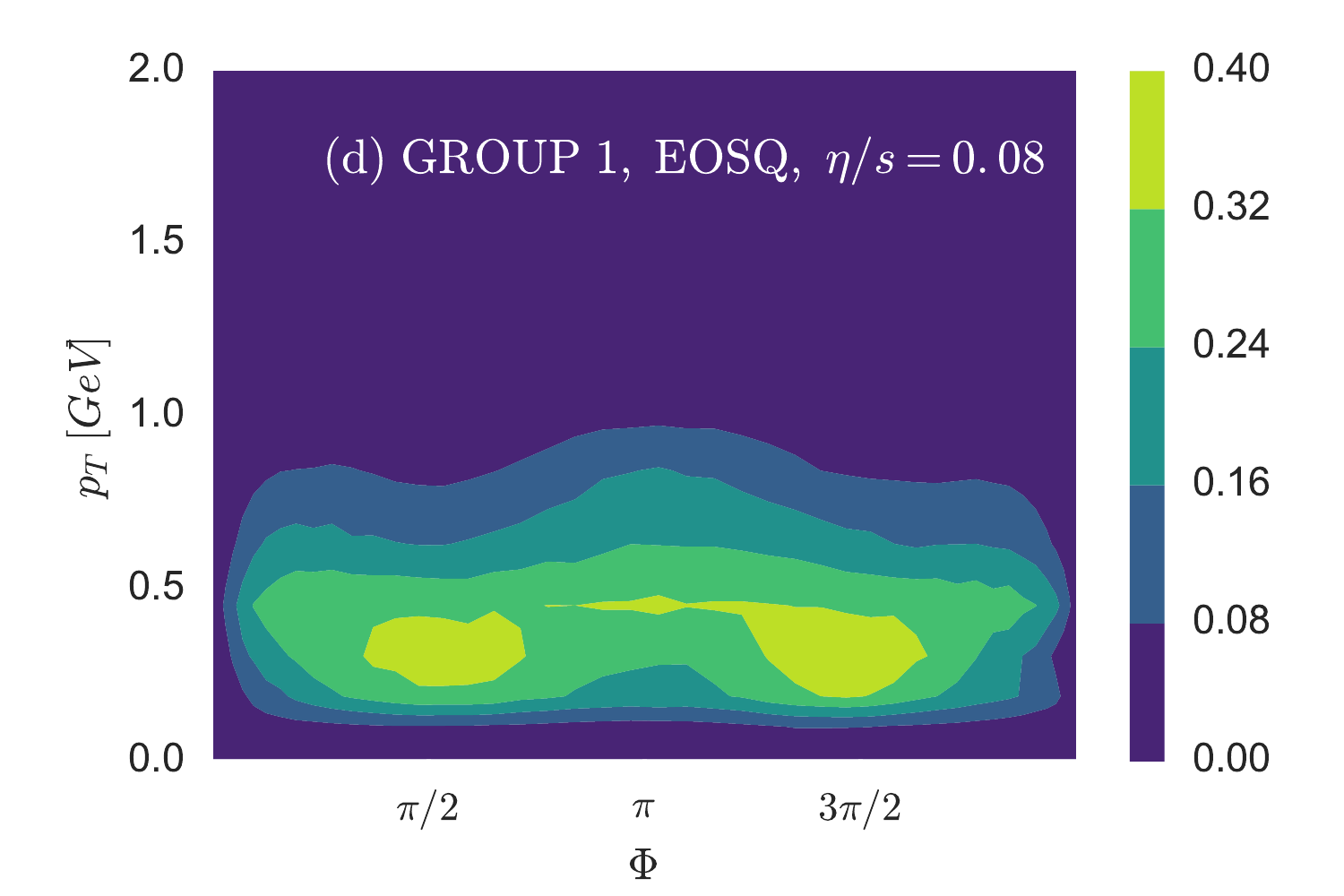}\includegraphics[width=0.33\textwidth, trim={0.5cm 0cm 0.5cm 0},clip]{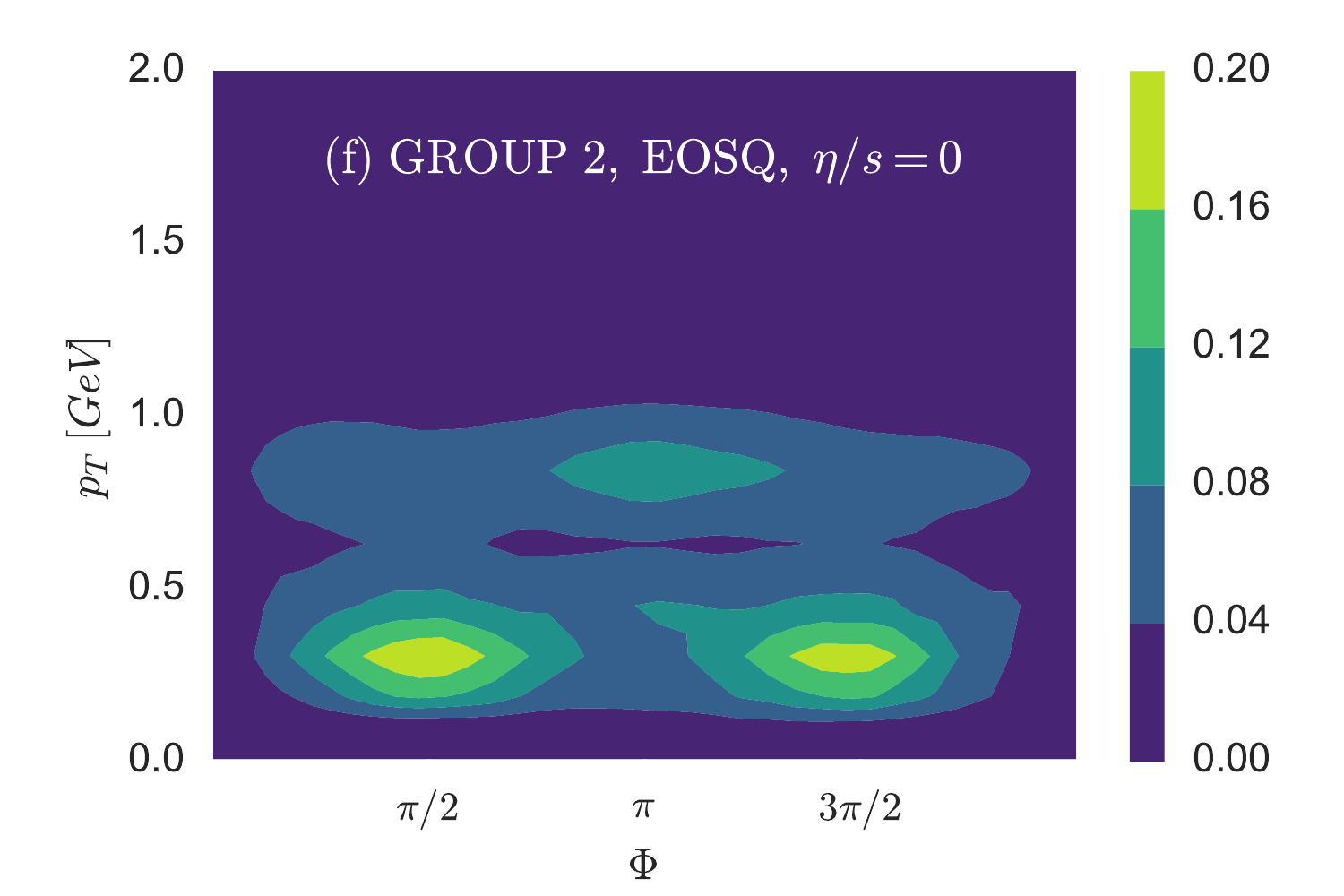}
  \caption{Importance maps of the $(p_T, \Phi)$ bins using the Prediction Difference Analysis method. The values of each bin are computed from event-average over about 800 events for each category (a) EOSL and (b) EOSQ with $\eta/s=0$, (c) EOSL and (d) EOSQ with $\eta/s=0.08$ from testing dataset GROUP 1, (e) EOSL and (f) EOSQ with $\eta/s=0$ from testing dataset GROUP 2.}
  \label{fig:weight_of_evidence}
\end{figure}

Shown in Fig.~\ref{fig:weight_of_evidence} are importance maps which illustrate the $(p_T, \Phi)$ dependence of the mean prediction difference averaged over 800 events for different model setups (initial conditions, PDE solver and model parameters), EoSs and values of the shear viscosity. For a given event, the mean prediction difference in each $(p_T, \Phi)$ bin is computed against 10 random reference events from the same dataset. Comparing different columns in the same row in Fig.~\ref{fig:weight_of_evidence},  we can see that importance maps vary slightly for different values of viscosity and model setups (Group 1: IEBE-VISHNU+MC-Glauber, Group 2: CLVics+IP-Glasma) for the same EoS. However, importance maps for EOSL in general have a distinctly narrower width in the $p_T$ range than that for EOSQ, independently of  the model setup and the value of viscosity \cite{Chaudhuri:2005ea}. This might be the important region of hidden features the network recognizes in classifying the EoS under each event.

\subsection{Conclusion}

The present method yields a novel perspective on identifying the nature of the QCD transition in heavy-ion collisions. With the help of deep CNNs and its well generalization performance, we firmly demonstrate that discriminative and traceable projections -- ``encoders'' --  from the QCD transition onto the final-state $\rho(p_T,\Phi)$ do exist in the complex and highly dynamical heavy-ion collisions, although these encoders may not be intuitive. The deep CNN provides a powerful and efficient ``decoder'' from which the EoS information can be extracted directly from the $\rho(p_T,\Phi)$. It is in this sense that the high-level representations, which help decoding the EoS information in the present method, act as an ``EoS-meter'' for the QCD matter created in heavy-ion collisions. The Prediction Difference Analysis method is employed to extract the most relevant features for the classification task, which may inspire phenomenological and experimental studies. Our study might provide a key to the success of the experimental determination of QCD EoS and search for the critical end point. Another intriguing application of our framework is to extract the QGP transport coefficients from heavy-ion collisions. The present method can be further improved by including hadronic rescattering and detector efficiency corrections.

\section{Method}
\label{s3}

The decisive ingredients for the success of hydrodynamic modeling of relativistic heavy-ion collisions are the bulk-matter EoS and the viscosity. In the study of the QCD transition in heavy-ion collisions, one of the holy-grail question is: how to reliably extract EoS and the nature of the QCD transition from the experimental data? The convolutional neural network (CNN)~\cite{cnn-12, 2014arXiv1409.1556S} is a powerful technique in tasks such as image and video recognition, natural language processing. Supervised training of the CNN with labeled $\rho(p_T,\phi)$ generated by CLVisc is tested with $\rho(p_T,\phi)$ generated by iEBE-VISHNU. The training and testing $\rho(p_T,\phi)$ can be regarded as numerical experimental data. Hence, analyzing real experimental data is possible with straightforward generalizations of the current prototype setup.

\subsection{Network architecture}

\begin{figure}[!htp]
  \centering
  \includegraphics[width=0.8\textwidth]{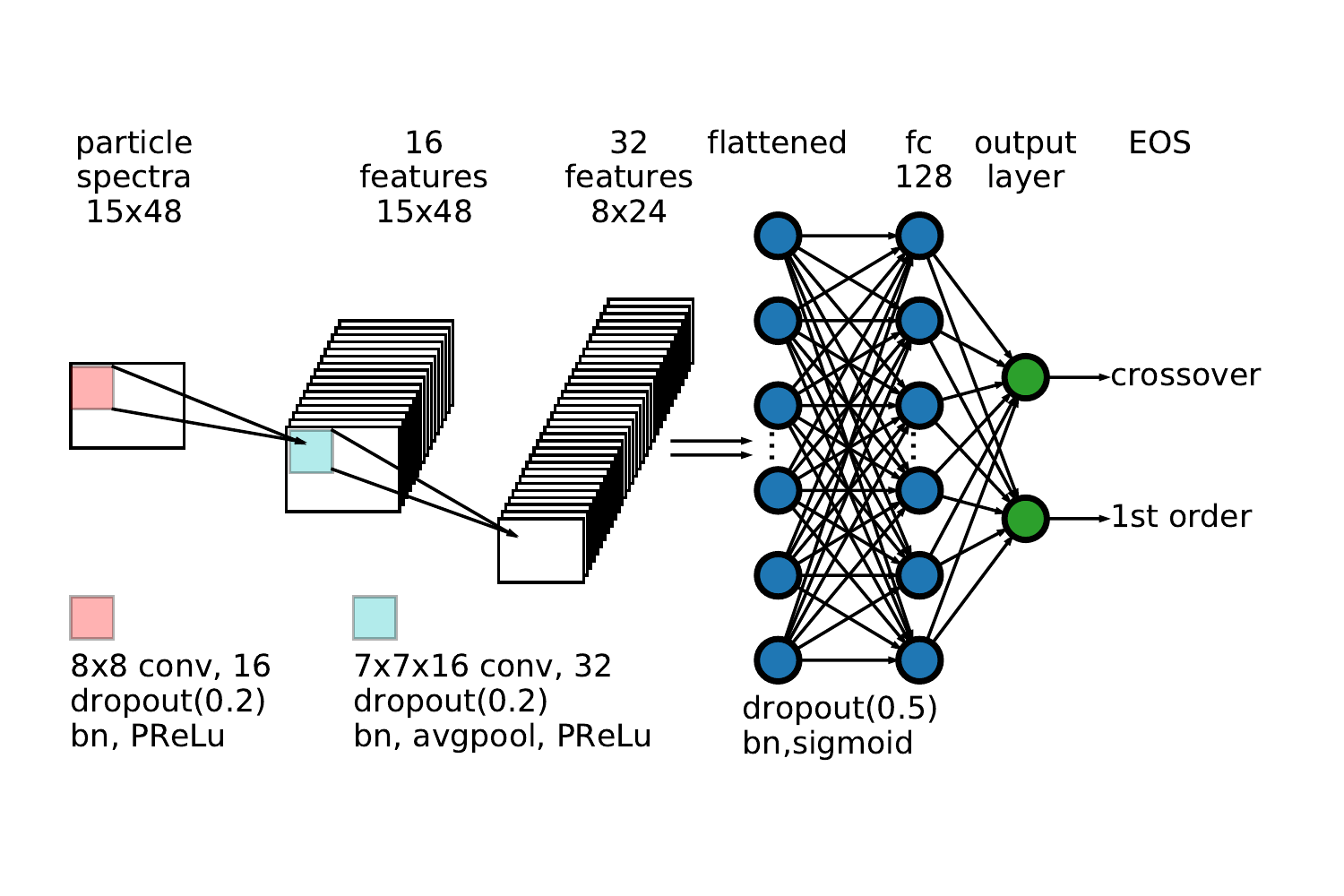}
  \caption{Our convolution neural network (CNN) architecture for identifying the QCD transition by using particle spectra with 15 transverse momentum $p_T$ bins and 48 azimuthal angle $\Phi$ bins.}
  \label{fig:cnn_eos}
\end{figure}

Our CNN architecture is shown in Fig.~\ref{fig:cnn_eos}. The input $\rho(p_T,\Phi)$ consists of 15 $p_T$-bins and 48 $\Phi$-bins. We use two convolutional layers each followed by batch normalization~\cite{2015arXiv150203167I}, dropout~\cite{Srivastava:2014} with a rate 0.2 and PReLU activation~\cite{2015arXiv150201852H}. These technical terms are briefly explained in the supplementary materials.
In the first convolutional layer, there are 16 filters of size $8\times8$ scanning through the input $\rho(p_T,\Phi)$ and creating 16 features of size $15\times48$. These features are further convoluted in the second convolutional layer that has $32$ filters of size $7\times7\times16$. The weight matrix of both convolutional layers are initialized with normal distribution and constrained with L2 regularization~\cite{L2}. In a convolutional layer, each neuron only locally connects to a small chunk of neurons in the previous layer by a convolution operation -- this is a key reason for the success of the CNN architecture. Dropout, batch normalization, PReLU and L2 regularization work together to prevent overfitting that may generate model-dependent features from the training dataset and thus hinder the generalizability of the method. The resulting 32 features of size $8\times24$ from the second convolutional layer are flattened and connected to a 128-neuron fully connected layer with batch normalization, dropout with rate 0.5 and sigmoid activation. The output layer is another fully connected layer with softmax activation and 2 neurons to indicate the type of the EoS. For multi-class classification, one may use more neurons in the output layer.

\subsection{Training and validation}

We use supervised learning to tackle this binary classification problem with the crossover case labeled by $(1, 0)$ and the first-order case labeled by $(0, 1)$. The difference between the true label and the predicted label from the two output neurons, quantified by cross entropy~\cite{crossEntropy}, serves as the loss function $l(\theta)$, where $\theta$ are the trainable parameters of the neural network. Training attempts to minimize the loss function by updating $\theta \rightarrow \theta - \delta \theta$. Here $\delta\theta=\alpha\ \partial l(\theta)/\partial \theta$ where $\alpha$ is the learning rate with initial value $0.0001$ and adaptively changed in AdaMax method~\cite{2014arXiv1412.6980K}.

We build the architecture using Keras~\cite{chollet2015keras} with a TensorFlow (r1.0)~\cite{tensorflow} backend and train the neural network with 2 NVIDIA GPUs K20m. The training dataset is fed into the network in batches with batch size empirically selected as 64. One traversal of all the batches in the training dataset is called one epoch. To accelerate the learning, the training dataset is reshuffled before each epoch. The neural network is trained with 500 epochs. Small fluctuations of validation accuracy saturated around $99\%$ are observed. The model parameters are saved to a new checkpoint whenever a smaller validation error is encountered.

\begin{figure}[!htp]
  \centering
  \includegraphics[width=0.8\textwidth]{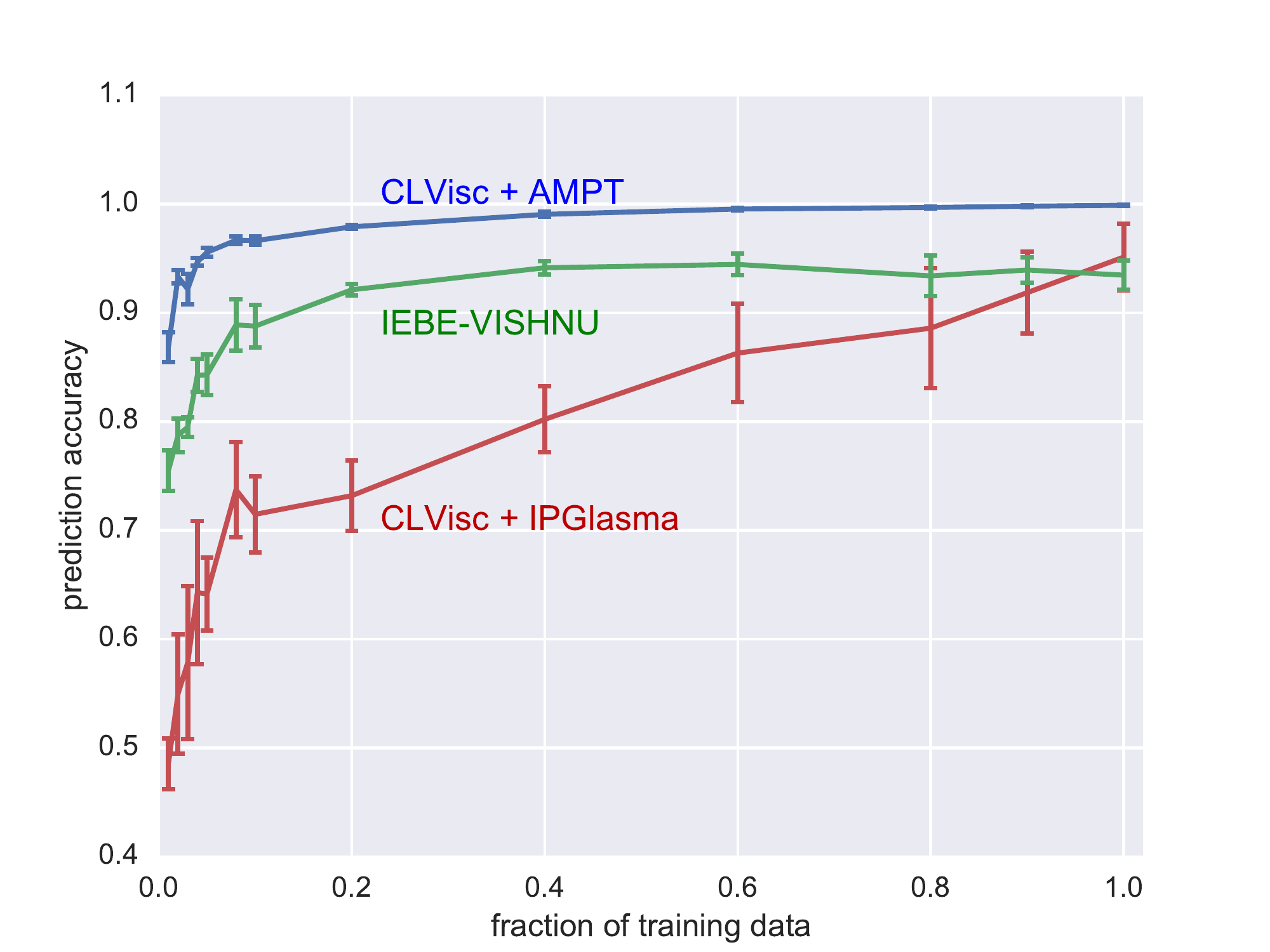}
  \caption{The mean and standard deviation of prediction accuracy in 10-fold cross validation tests when different fractions of the training data is used to train the network.}
  \label{fig:cvs_vs_size}
\end{figure}

The k-fold stratified cross validation is employed to estimate the model uncertainties. The training dataset is randomly shuffled and  split into $k$ equal folds with each fold contains equal number of two types of training data. One of these $k$ folds is used for validation while the other $k-1$ folds are used for training. Finally $k$ models (according to $k$ pairs of (training, validation) partitioning) are trained to get the mean prediction accuracy and standard deviation. As shown in Fig.~\ref{fig:cvs_vs_size}, the prediction accuracy approaches $99\%$ with negligible uncertainty for  testing on CLVisc+AMPT (same data generator as training), using less than $50\%$ of the training data. While for the testing on IEBE-VISHNU + MC-Glauber (testing GROUP 1) and CLVisc + IP-Glasma (testing GROUP 2), the prediction accuracy increases as one increases the size of the training dataset, which is in line with the practical expectation that more training data could boost the network's performance. With the full training data, we get on average  a larger than $95\%$ prediction accuracy, which is a very positive manifestation of the generalization capability of our deep CNN.

For the network settings, most of the parameters are introduced in the fully connected layers. In an alternative model, we add 2 more convolutional layers with filter size (3, 3) and subsequent average pooling layers to reduce the number of neurons in the flatten layer and also in the first fully connected layer, which helps to reduce the total number of parameters by a factor of $10$. This deeper neural network produces similar prediction accuracy and model uncertainty in a 10-fold cross validation tests. 

\section{Supplementary Material}

\begin{figure}[!htp]
  \centering
  \includegraphics[width=\textwidth, trim=0 5cm 0 8cm, clip]{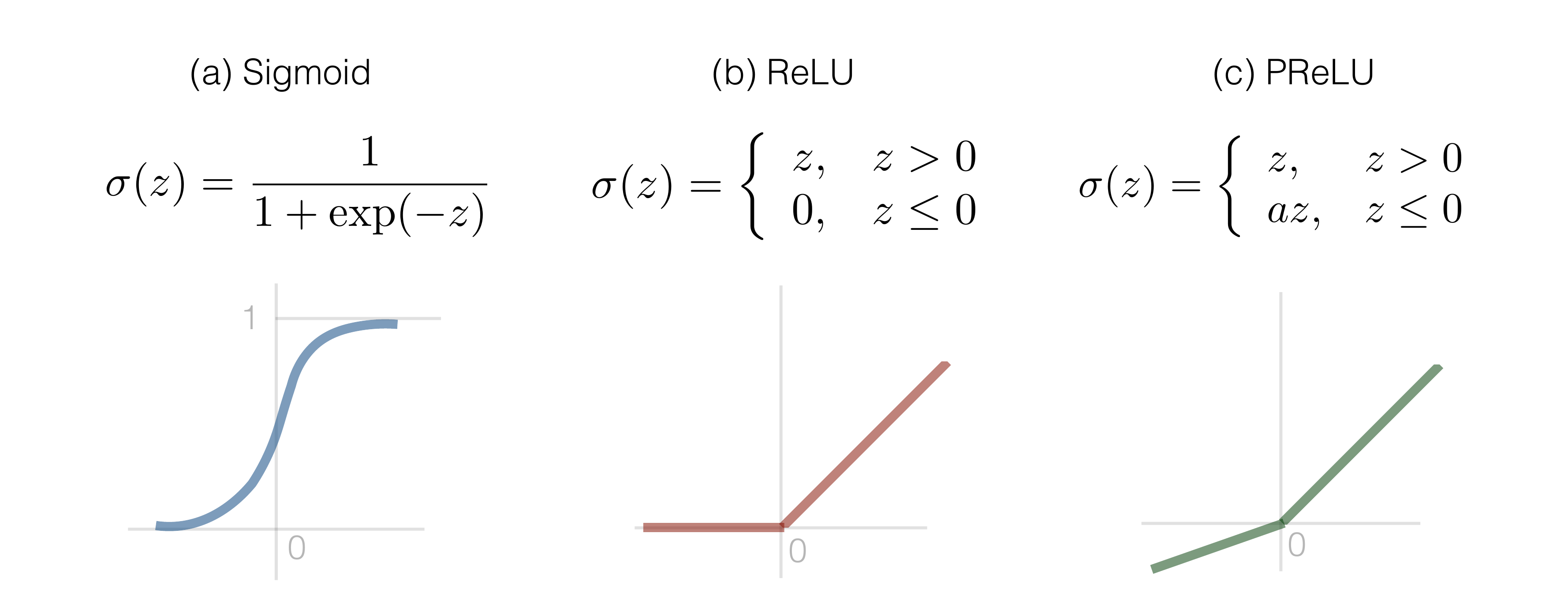}
  \caption{(a) Sigmoid, the logistic function which has an `S' shaped curve (b) ReLU, rectified linear unit that activates the neuron when $z>0$ and (c) PReLU parametric rectified linear unit that additionally activates leaky neurons at $z<0$ with learnable  parameter $a$.}
  \label{fig:activation}
\end{figure}

\textit{Feedforward neural network} 
learns one target function $\mathbf{x}: f(\mathbf{x}, \theta) \rightarrow \mathbf{y}$ that maps  the input vector $\mathbf{x}$ to output vector $\mathbf{y}$ with parameter $\theta$. Elements of $\mathbf{x}$ and $\mathbf{y}$ form the neurons in the input and output layers respectively.  In-between there can be multiple hidden layers with the numbers of neurons as hyper-parameters. The connections between two layers form a trainable weight matrix $W$. Each layer (except the input layer) learns representations of its previous layer through firstly a linear operation $\mathbf{z}=\mathbf{x} W + \mathbf{b}$ and then use it as the argument of an activation function $\sigma(\mathbf{z})$. The linear operation can perform various operations, such as scaling, rotating, boosting, increasing or decreasing dimensions, on the vector $\mathbf{x}$, with the bias $\mathbf{b}$ a trainable parameter. $\sigma(\mathbf{z})$ activates the neurons of the present layer with their values and computes the correlations between the neurons of the previous layer. For classification network, softmax activation function $\sigma({\bf z})_j=exp(z_j)/\sum_{k=1}^{K}exp(z_k)$ is usually used in the final layer to compute the probability of each category. By stacking with multiple hidden layers, the deep neural network may learn high-level representations that can be classified or interpreted easily. The activation functions used in our study are shown in Fig.~\ref{fig:activation}.

\textit{Loss function}
$l(\theta)$ is the difference between the true value $\mathbf{y}$ (from the input of supervised learning) and the predicted value  $\mathbf{\hat{y}} = f(\mathbf{x}, \theta)$ by the neural network in a forward pass. The simplest loss function is the mean square error $l(\theta) = \sum_i(\hat{y}_i - y_i)^2$. In this paper we use the cross entropy loss function from information theory,
\begin{equation}
l(\theta) = -\frac{1}{N} \sum_{i=1}^{N} \left[   y_i \log \hat{y}_i + (1 - y_i) \log (1 - \hat{y}_i) \right]
\label{eq:cross_entropy}
\end{equation}
With L1 or L2 regularizations, the loss function receives another term used to constrain the values of $\theta$ from going wildly,
\begin{eqnarray}
L1:\ l(\theta) &=&  l(\theta) + \lambda || \theta ||_1  \\
L2:\ l(\theta) &=&  l(\theta) + \lambda || \theta ||_2^2
\label{eq:l1_l2}
\end{eqnarray}
where $\lambda$ is the regularization strength, $||\theta ||_p \equiv \left(\sum_j^n |\theta_j|^p \right)^{1/p}$ is the $p$-norm of the parameters $\theta=(\theta_1, \theta_2, ..., \theta_n)$. Larger $\lambda$ leads to smaller $\theta$, especially for high orders in the target function,  which increases the generalizability of the neural network.

\textit{Back propagation}
indicates the gradients of the loss function in parameter space propagate in the backward direction of a neural network in order to update $\theta$. For example, in the stochastic gradient decent (SGD) method, $\theta$ is updated with fixed learning rate $\epsilon$
\begin{equation}
\theta^{'} = \theta - \epsilon \frac{\partial l(\theta)}{\partial \theta}
\label{eq:cross_entropy}
\end{equation}
In practice we train the network in batches, where $\theta$ is updated once for all the samples in one batch,
\begin{equation}
\theta^{'} = \theta - \frac{\epsilon}{m} \sum_{i=1}^{m} \frac{\partial l_i(\theta)}{\partial \theta}
\label{eq:cross_entropy}
\end{equation}
where $m$ is the batchsize, $l_i$ is the loss given by the $i$th training sample in a batch. In our study, we use the AdaMax method \cite{2014arXiv1412.6980K}, which computes adaptive learning rates for different parameters based on estimating the first and second moments of the gradients. We initially set the learning rate as $\alpha=10^{-4}$ and keep the other parameters the same as in \cite{2014arXiv1412.6980K}.

\textit{Batch normalization} 
solves the internal covariate shift problem, which is a common issue in DL that hinders the learning efficiency~\cite{2015arXiv150203167I}. Using the batch mean $\mu_B = \frac{1}{m} \sum_{i=1}^{m} x_i$ and batch variance  $\sigma_B^2 = \frac{1}{m} \sum_{i=1}^{m} (x_i - \mu_B)^2$, the input vector $\mathbf{x}$ is normalized as $\hat{x}_i = \frac{x_i - \mu_B}{\sqrt{\sigma_B^2 + \epsilon}}$ that has mean 0 and variance 1, with $\epsilon$ a small number preventing divergence. The $\hat{\mathbf{x}}$ is further scaled and shifted by $\gamma \hat{\mathbf{x}} + \beta$ before going to the next layer, where $\gamma$ and $\beta$ are trainable parameters. Note that during the testing, population mean and variance of the training dataset are used.

\textit{Dropout} 
is a regularization technique that reduces overfitting  by randomly discarding a fraction of neurons (features) and all their associated connections to prevent co-adaption~\cite{dropout} of neurons for each training sample .

\textit{Prediction Difference Analysis} 
is a method to visualize the difference between the log-odds of the prediction probability $p(y | \rho)$ and $p(y | \rho_{ \backslash i})$, where $y$ is the class value, $\rho$ is the real image and $\rho_{\backslash i}$ is the imperfect image without the knowledge of the $i$th pixel. The prediction difference is dubbed as \textit{weight of evidence}~\cite{robnik2008explaining,Luisa:2017},
\begin{equation}
\mathrm{WE}_i(y | \rho) = \log_2 \left( odds(y|\rho) \right) - \log_2 \left( odds(y|\rho_{\backslash i}) \right)\,,
\end{equation}
where $odds(z) = p(z) / (1 - p(z))$ is used to symmetrize $\log_2 p$ and $-\log_2(1-p)$,  with Laplace correction $p \leftarrow (p n+ 1)/(n + m)$ to avoid zero probability, where $n$ is the number of training instances and $m$ is the number of classes. The $p(y | \rho_{ \backslash i})$ is approximated by,
\begin{equation}
p(y | \rho_{\backslash i}) \approx \sum_{s}^{m_i} p(\rho_i) p(y | \rho \leftarrow \rho_i = a_s) \,,
\end{equation}
with the $i$th pixel replaced with all the possible values $a_s$ weighted by its value probability. The importance map is given by the mean weight of evidence over many events that have the same class label.

\begin{acknowledgments}

L.G.P. and H.P. acknowledge funding of a Helmholtz Young Investigator Group VH-NG-822 from the Helmholtz Association and the GSI Helmholtzzentrum f\"ur Schwerionenforschung (GSI). K.Z. and N.S. acknowledge the support from GSI. H.St. acknowledges the support through the Judah M. Eisenberg Laureatus Chair at Goethe University. X.N.W was supported in part by NSFC under the Grant No. 11521064, by MOST of China under Grant No. 2014DFG02050, by the Major State Basic Research Development Program (MSBRD) in China under the Grant No. 2015CB856902and  by U.S. DOE under Contract No. DE-AC02-05CH11231. This work was supported in part by the Helmholtz International Center for the Facility for Antiproton and Ion Research (HIC for FAIR) within the framework of the Landes-Offensive zur Entwicklung Wissenschaftlich-Oekonomischer Exzellenz (LOEWE) program launched by the State of Hesse. The computations were done in the Green-Cube GPU cluster LCSC at GSI, the Loewe-CSC at Goethe University, and the GPU cluster at Central China Normal University.

\end{acknowledgments}

{\bf Author Contributions:}
L.G.Pang contributed to the idea, the training and the second testing dataset, the neural network construction for training/testing and the manuscript preparation; K.Zhou contributed to the idea, the first testing dataset, intensive discussions on neural network structures, physical explanations of the results and the manuscript edition; N.Su contributed to intensive discussions on neural network structures, physical explanations of the results and the manuscript edition; H.Petersen, H. Stocker and X.N.Wang contributed to the computing resources,  physical insights and manuscript editions.

{\bf Corresponding author:}
Correspondence to L.G. Pang, K. Zhou and N. Su. Emails:
\{pang, zhou, nansu\}@fias.uni-frankfurt.de

{\bf Competing interests:}
The authors declare no competing financial interests.


\begin{thebibliography}{99}
  
\bibitem{dl1}
  J.~Schmidhuber,
  Neural Netw.\ {\bf 61}, 85 (2015).
  
\bibitem{dl2}
  Y.~LeCun, Y.~Bengio, and G.~Hinton,
  Nature {\bf 521}, 436 (2015).
  
\bibitem{Baldi:2014kfa}
  P.~Baldi, P.~Sadowski, and D.~Whiteson,
  Nature Commun.\  {\bf 5}, 4308 (2014).
  
\bibitem{Baldi:2014pta}
  P.~Baldi, P.~Sadowski, and D.~Whiteson,
  Phys.\ Rev.\ Lett.\  {\bf 114}, 111801 (2015).
  
\bibitem{Searcy:2015apa}
  J.~Searcy, L.~Huang, M.~A.~Pleier, and J.~Zhu,
  Phys.\ Rev.\ D {\bf 93}, 094033 (2016).
  
\bibitem{Barnard:2016qma}
  J.~Barnard, E.~N.~Dawe, M.~J.~Dolan, and N.~Rajcic,
  Phys.\ Rev.\ D {\bf 95}, 014018 (2017).
  
\bibitem{Moult:2016cvt}
  I.~Moult, L.~Necib, and J.~Thaler,
  J.\ High Energy Phys.\ {\bf 12}, 153 (2016).

\bibitem{Utama:2016tcl}
  R.~Utama, W.~C.~Chen, and J.~Piekarewicz,
  J.\ Phys.\ G {\bf 43}, 114002 (2016).
  
\bibitem{rg&dl}
  P.~Mehta and D.~J.~Schwab,
  arXiv:1410.3831 [stat.ML].

\bibitem{ml-phases}
  J.~Carrasquilla and R.~G.~Melko,
  Nat.\ Phys.\ \url{http://dx.doi.org/10.1038/nphys4035} (2017).

\bibitem{quantum-many-body-ann}
  G.~Carleo and M.~Troyer,
  Science {\bf 355}, 602 (2017).
  
\bibitem{eos-bm}  
  G.~Torlai and R.~G.~Melko,
  Phys.\ Rev.\ B {\bf 94}, 165134 (2016).
  
\bibitem{sign}
  P.~Broecker, J.~Carrasquilla, R.~G.~Melko, and S.~Trebst,
  arXiv:1608.07848 [cond-mat.str-el].
  
\bibitem{str-f}
  K.~Ch'ng, J.~Carrasquilla, R.~G.~Melko, and E.~Khatami,
  arXiv:1609.02552 [cond-mat.str-el].
  
\bibitem{Stoecker:1986ci}
  H.~St\"ocker and W.~Greiner,
  Phys.\ Rept.\  {\bf 137}, 277 (1986).
  
\bibitem{Stephanov:2007fk}
  M.~A.~Stephanov,
  PoS LAT {\bf 2006} (2006) 024.
  
\bibitem{Fukushima:2010bq}
  K.~Fukushima and T.~Hatsuda,
  Rept.\ Prog.\ Phys.\  {\bf 74}, 014001 (2011).

\bibitem{Luzum:2008cw} 
  M.~Luzum and P.~Romatschke,
  Phys.\ Rev.\ C {\bf 78}, 034915 (2008).
 
\bibitem{Heinz:2009xj}
  U.~W.~Heinz,
  Landolt-Bornstein {\bf 23}, 240 (2010).
  
\bibitem{Romatschke:2009im}
  P.~Romatschke,
  Int.\ J.\ Mod.\ Phys.\ E {\bf 19}, 1 (2010).
  
\bibitem{Teaney:2009qa}
  D.~A.~Teaney,
  arXiv:0905.2433 [nucl-th].
  
\bibitem{Gale:2013da}
  C.~Gale, S.~Jeon, and B.~Schenke,
  Int.\ J.\ Mod.\ Phys.\ A {\bf 28}, 1340011 (2013).
  
\bibitem{Strickland:2014pga}
  M.~Strickland,
  Acta Phys.\ Polon.\ B {\bf 45}, 2355 (2014).
    
\bibitem{Pang:2012he} 
  L.~G.~Pang, Q.~Wang, and X.~N.~Wang,
  Phys.\ Rev.\ C {\bf 86}, 024911 (2012).
  
\bibitem{Pang:2014ipa} 
  L.~G.~Pang, Y.~Hatta, X.~N.~Wang, and B.~W.~Xiao,
  Phys.\ Rev.\ D {\bf 91}, 074027 (2015).
  
\bibitem{Huovinen:2009yb} 
  P.~Huovinen and P.~Petreczky,
  Nucl.\ Phys.\ A {\bf 837}, 26 (2010).
  
\bibitem{Sollfrank:1996hd} 
  J.~Sollfrank, P.~Huovinen, M.~Kataja, P.~V.~Ruuskanen, M.~Prakash, and R.~Venugopalan,
  Phys.\ Rev.\ C {\bf 55}, 392 (1997).
  
\bibitem{Pratt:2015zsa} 
  S.~Pratt, E.~Sangaline, P.~Sorensen, and H.~Wang,
  Phys.\ Rev.\ Lett.\  {\bf 114}, 202301 (2015)
  
\bibitem{Bernhard:2016tnd} 
  J.~E.~Bernhard, J.~S.~Moreland, S.~A.~Bass, J.~Liu, and U.~Heinz,
  Phys.\ Rev.\ C {\bf 94}, 024907 (2016)

\bibitem{Shen:2014vra} 
  C.~Shen, Z.~Qiu, H.~Song, J.~Bernhard, S.~Bass, and U.~Heinz,
  Comput.\ Phys.\ Commun.\  {\bf 199}, 61 (2016).
  
\bibitem{Stoecker:2004qu}
  H.~St\"ocker,
  Nucl.\ Phys.\ A {\bf 750}, 121 (2005).
  
\bibitem{Lin:2004en} 
  Z.~W.~Lin, C.~M.~Ko, B.~A.~Li, B.~Zhang, and S.~Pal,
  Phys.\ Rev.\ C {\bf 72}, 064901 (2005).

\bibitem{Gale:2012rq} 
  C.~Gale, S.~Jeon, B.~Schenke, P.~Tribedy, and R.~Venugopalan,
  Phys.\ Rev.\ Lett.\  {\bf 110}, 012302 (2013).

\bibitem{robnik2008explaining}
  M.~Robnik-Sikonja and I.~Kononenko,
  Knowledge and Data Engineering, IEEE Transactions on, 20(5):589-600, (2008).

\bibitem{Luisa:2017}
  L.~M.~Zintgraf, T.~S.~Cohen, T.~Adel, and M.~Welling,
  arXiv:1702.04595 [cs.CV].

\bibitem{Chaudhuri:2005ea} 
  A.~K.~Chaudhuri and U.~W.~Heinz,
  J.\ Phys.\ Conf.\ Ser.\  {\bf 50}, 251 (2006)
  
\bibitem{cnn-12}
  A.~Krizhevsky, I.~Sutskever, and G.~E.~Hinton,
  Advances in Neural Information Processing Systems 25 (NIPS 2012).
  
\bibitem{2014arXiv1409.1556S} 
  K.~Simonyan and A.~Zisserman, 
  arXiv:1409.1556 [cs.CV].
  
\bibitem{2015arXiv150203167I} 
  S.~Ioffe and C.~Szegedy, 
  arXiv:1502.03167 [cs.LG].

\bibitem{Srivastava:2014} 
  N.~Srivastava {\it et al.},
  J.\ Mach.\ Learn.\ Res.\ {\bf 15}, 1929 (2014).

\bibitem{2015arXiv150201852H} 
  K.~He, X,~Zhang, S.~Ren, and J.~Sun, 
  arXiv:1502.01852 [cs.CV].

\bibitem{L2} 
  A.~Y.~Ng,
  Proceedings of the 21st International Conference on Machine Learning, Banff, Canada, 2004.

\bibitem{crossEntropy}
  S.~Kullback and R.~A.~Leibler,
  Ann.\ Math.\ Statist.\ {\bf 22}, 79 (1951).

\bibitem{2014arXiv1412.6980K} 
  D.~Kingma and J.~Ba, 
  arXiv:1412.6980 [cs.LG].

\bibitem{chollet2015keras} 
  F.~Chollet, 
  \url{https://github.com/fchollet/keras}.

\bibitem{tensorflow} 
  M.~Abadi, {\it et al.}, 
  arXiv:1603.04467 [cs.DC], 
  \url{http://tensorflow.org/}.

\bibitem{dropout} G.~E.~Hinton, N.~Srivastava, A.~Krizhevsky, I.~Sutskever, and R.~R.~Salakhutdinov,
arXiv:1207.0580 [cs.NE].


\end{thebibliography}
\end{document}